\documentstyle[12pt]{article}
\makeatletter
\renewcommand{\section}{\@startsection {section}{1}{\z@}
{-3.5ex plus -1ex minus -.2ex}{2.3ex plus .2ex}{\normalsize\bf}}
\renewcommand{\subsection}{\@startsection{subsection}{2}{\z@}
{-3.25ex plus -1ex minus -.2ex}{1.5ex plus .2ex}{\normalsize\it}}

\def\abstract{\if@twocolumn
\section*{\abstractname}
\else \small
\quotation
\fi}
\def\endabstract{\if@twocolumn\else\endquotation\fi}
\renewcommand{\@makefnmark}{\hbox{\mathsurround=0pt
$^\dagger$}}
\renewcommand{\@makefntext}[1]{\parindent=1em\noindent
\hbox to 1.8em{\hss$^\dagger$}#1}
\def\thebibliography#1{\section*{\refname\@mkboth
 {\uppercase{\refname}}{\uppercase{\refname}}}\list
 {[\arabic{enumi}]}{\settowidth\labelwidth{[#1]}\leftmargin\labelwidth
 \advance\leftmargin\labelsep\parsep=0em\itemsep=0em
 \usecounter{enumi}}
 \def\newblock{\hskip .11em plus .33em minus .07em}
 \sloppy\clubpenalty4000\widowpenalty4000
 \sfcode`\.=1000\relax}
\makeatother
\input tcilatex
\QQQ{Language}{
American English
}

\begin{document}

\begin{center}
{\normalsize {\bf QUANTITATIVE DESCRIPTION OF THERMODYNAMICS OF LAYERED
MAGNETS IN A WIDE TEMPERATURE REGION}} \\\bigskip
V.Yu. Irkhin\footnote{%
Corresponding author. Fax: +7 (3432) 74 52 44; e-mail:
Valentin.Irkhin@imp.uran.ru} and A.A. Katanin\medskip \\{\small {\it %
Institute of Metal Physics, 620219 Ekaterinburg, Russia}}
\end{center}

\begin{abstract}
The thermodynamics of layered antiferro- and ferromagnets with a weak
interlayer coupling and/or easy-axis anisotropy is investigated. A crossover
from an isotropic 2D-like to 3D Heisenberg (or 2D Ising) regime is discussed
within the renormalization group (RG) analysis. Analytical results for the
the (sublattice) magnetization and the ordering temperature are derived are
obtained in different regimes. Numerical calculations on the base of the
equations obtained yield a good agreement with experimental data on La$_2$CuO%
$_4$ and layered perovskites. Corresponding results for the
Kosterlitz-Thouless and Curie (N\'eel) temperatures in the case of the
easy-plane anisotropy are derived.\bigskip \\\noindent

{\em PACS:\/} 75.10 Jm, 75.30.Gw, 75.70.Ak

\noindent
{\em Keywords:\/} Layered magnetic systems, renormalization group, $1/N$
expansion.
\end{abstract}

The problem of layered magnetic systems is of interest both from theoretical
and practical point of view \cite{Joungh}. Here belong, e.g.,
quasi-two-dimensional (quasi-2D) perovskites, ferromagnetic monolayers and
ultrathin films. We consider the Heisenberg model with small parameters of
the interlayer coupling $\alpha =J^{\prime }/J$ and easy-axis anisotropy $%
\eta =1-J^z/J^{x,y}$ ($J$ is the in-plane exchange parameter). This case
permits a regular consideration since the magnetic transition temperature is
small, $T_M\ll |J|S^2,$ which is reminiscent of weak itinerant magnets.

At not too low temperatures $T$ the standard spin-wave theory (SWT) is
insufficient to describe correctly thermodynamics of such systems. Somewhat
better results can be obtained within the self-consistent theory (SSWT) \cite
{Takahashi,OurSSWT}, which takes into account the temperature
renormalization of $\alpha $ and $\eta .$ However, the values of the
ordering temperature in SSWT are still too high in comparison with
experimental ones, and the critical behavior is quite incorrect.

To improve radically SSWT, the summation of leading contributions in all
orders of perturbation theory should be performed. To this end we use the
renormalization group (RG) approach similar to that of Ref. \cite
{Chakraverty}. This approach is valid outside the critical region which is
very narrow for layered systems. The same results can be obtained by direct
summation of RPA-type corrections to spin-wave interaction vertex.

The result for the relative (sublattice) magnetization $\overline{\sigma }%
_r\equiv \overline{S}/\overline{S}(T=0)$ reads \cite{OurSSWT,OurRG}
\begin{equation}
\overline{\sigma }_r=1-\frac T{4\pi \rho _s}\left[ \ln \frac{\Gamma (T)}{%
\Delta (f_T,\alpha _T)}+2\ln (1/\overline{\sigma }_r)+2(1-\overline{\sigma }%
_r)+\Phi \left( \frac T{4\pi \rho _s\overline{\sigma }_r}\right) \right]
\label{fRG}
\end{equation}
with $\Delta (f,\alpha )=f+\alpha +\sqrt{f^2+2\alpha f}$, $f=4\eta ,$ other
quantities are given in the Table 1:
\[
\begin{tabular}{||lll|ll||}
\hline\hline
\multicolumn{1}{||l|}{} & \multicolumn{1}{l|}{$\Gamma $} &
\multicolumn{1}{l|}{$\rho _s$} & \multicolumn{1}{|l|}{$f_r$} & $\alpha _r$
\\ \hline
\multicolumn{1}{||l|}{AFM, quantum regime ($T\ll |J|S$)} &
\multicolumn{1}{l|}{$T^2/c^2$} & \multicolumn{1}{l|}{$\gamma |J|S\overline{S}%
_0$} & \multicolumn{1}{|l|}{$f\overline{S}_0^2/S^2$} & $\alpha \overline{S}%
_0/S$ \\
\multicolumn{1}{||l|}{FM, $\;\;$quantum regime ($T\ll JS$)} &
\multicolumn{1}{l|}{$T/JS$} & \multicolumn{1}{l|}{$JS^2$} &
\multicolumn{1}{|l|}{$f$} & $\alpha $ \\
\multicolumn{1}{||l|}{FM,AFM, classical regime ($T\gg |J|S$)} &
\multicolumn{1}{l|}{$32$} & \multicolumn{1}{l|}{$JS^2Z_{L1}$} &
\multicolumn{1}{|l|}{$fZ_{L2}$} & $\alpha Z_{L3}$ \\ \hline\hline
\end{tabular}
\]
where $c=\sqrt{8}|J|\gamma S$ is the spin-wave velocity, $\gamma \simeq
1+0.078/S$ is the renormalization parameter for intralayer coupling, $%
Z_{L1}=Z_{L2}=Z_{L3}=1-T/8\pi |J|S^2$. The inequality $\Gamma \gg \Delta $
should to be satisfied for the validity of (\ref{fRG}). Note that the
classical regime is realized only for very large $S$. The quantities $f_T$
and $\alpha _T$ in (\ref{fRG}) are the temperature-renormalized values of
interlayer coupling and anisotropy parameters, and for $T\ll 4\pi \rho _s%
\overline{\sigma }_r$ (i.e. beyond the critical region, see below)
\begin{equation}
f_T/f_r=(\alpha _T/\alpha _r)^2=\overline{\sigma }_r^2  \label{at}
\end{equation}
Thus the anisotropy and interlayer coupling are strongly renormalized with
temperature which should be taken into account when treating the
experimental data. In the quantum regime, the parameters $\alpha _r$ and $f_r
$ are ground-state (quantum-renormalized) anisotropy and interlayer coupling
(for ferromagnetic case the ground-state renormalizations are absent). In
the classical regime, $f_r$ and $\alpha _r$ are `lattice-renormalized'
parameters of anisotropy and interlayer coupling, i.e. the corresponding
parameters of the continuum model with the same thermodynamic properties as
the original lattice model.

Three temperature regimes can be distinguished.

\noindent (i) low temperatures, $T\ll T_M\sim 2\pi |J|S^2/\ln (\Gamma
/\Delta ).$ The analysis of non-uniform susceptibility shows that the
excitations in the whole Brillouin zone have spin-wave nature. Only first
term in the square brackets in (\ref{fRG}) is to be taken into account and
the magnetization also demonstrates the spin-wave behavior.

\noindent (ii) intermediate temperatures, $(\overline{S}/S)/\ln (\Gamma
/\Delta )\ll T/2\pi |J|S^2\ll \overline{S}/S$ ($T$ is of the same order as $%
T_M$). Close to the center of the Brillouin zone the excitations still have
the spin-wave nature, while for large enough momenta they have non-spin-wave
character. Only in-plane (two-dimensional) fluctuations are important in
this regime. All the terms in (\ref{fRG}), except for the last, are
important, which leads to significant modification of the dependence $%
\overline{\sigma }_r(T)$ in comparison with SWT.

\noindent (iii) critical region, $T/(2\pi |J|S^2)\gg \overline{S}/S\;$($%
1-T/T_M\ll 1$). In this regime the spin-wave excitations are present only
for momenta $q^2\ll \Delta $ (hydrodynamic region) whereas at all other $q$
excitations have non-spin-wave character. The thermodynamics in this case is
determined by 3D (or Ising-like) fluctuations. The contribution of $\Phi _{%
\text{a}}$ is of the same order as other terms in the square brackets of (%
\ref{fRG}) and the RG approach is unable to describe the thermodynamics in
this regime. The values of critical exponents can be corrected in comparison
with SSWT with the use of the $1/N$ expansion. For the isotropic quasi-2D
antiferromagnet we obtain \cite{Our1/N}
\begin{equation}
\overline{\sigma }_r^2=\left[ \frac{T_N}{4\pi |J|S\overline{S}_0\gamma }%
\right] ^{1-\beta _3}\left[ \frac 1{1-A_0}\left( 1-\frac T{T_N}\right)
\right] ^{2\beta _3}  \label{MagnCr1}
\end{equation}
with $A_0\simeq 0.9635$ and $\beta _3=(1-8/\pi ^2N)/2\simeq 0.36.$

Up to some (unknown) constant $C(f/\alpha )$ we have for the transition
temperature the result
\begin{equation}
1=\frac{T_M}{4\pi \rho _s}\left[ \ln \frac{2\Gamma (T_M)}{\Delta (f_t,\alpha
_t)}+2\ln \frac{4\pi \rho _s}{T_M}+C(f/\alpha )\right] .  \label{I}
\end{equation}
Note that all the logarithmic terms are included in (\ref{I}) and $C$ gives
only a small contribution to this result. For the quantum antiferromagnets
the value of $C(\infty )$ can be calculated by the $1/N$ expansion \cite
{Our1/N}, $C(\infty )\simeq -0.0660.$ The value of $C(0)$ for the same case
can be deduced from experimental data on layered magnetic compounds \cite
{OurSSWT}, $C(0)\simeq -0.7.$

For practical purposes, simple interpolation expressions for the functions $%
\Phi (x),$ which permit to describe the crossover temperature region, are
useful. We obtain at $x<1$ \cite{OurSSWT}:
\begin{eqnarray}
\left. \Phi (x)\right| _{\alpha =0} &=&\frac x{\sqrt{x^2+1}}\left[ C(\infty
)-2+8\ln 2\right] ,  \nonumber \\
\left. \Phi (x)\right| _{f=0} &=&\frac x{\sqrt{x^2+1}}\left[ C(0)-1+3\ln
3\right] .  \label{FF}
\end{eqnarray}

Numerical calculations with the use of the equations obtained yield a good
agreement with experimental data on the layered perovskites (see Figs. 1-2
and the Table 2), and the Monte-Carlo results for the anisotropic classical
systems.

\[
\begin{tabular}{||l|c|c|c|cc||}
\hline\hline
\multicolumn{1}{||l|}{Compound} & \multicolumn{1}{c|}{La$_2$CuO$_4$} & 
\multicolumn{1}{c|}{K$_2$NiF$_4$} & \multicolumn{1}{c|}{Rb$_2$NiF$_4$} & 
\multicolumn{1}{|c|}{K$_2$MnF$_4$} & CrBr$_3$ \\ \hline\hline
\multicolumn{1}{||l|}{$T_M^{\text{SSWT}},$K} & \multicolumn{1}{c|}{527} & 
\multicolumn{1}{c|}{125} & \multicolumn{1}{c|}{118} & \multicolumn{1}{|c|}{
52.1} & 51.2 \\ 
\multicolumn{1}{||l|}{$T_M^{\text{RG}},$K} & 343 & 97.0 & 95.0 & 
\multicolumn{1}{|c|}{42.7} & 39.0 \\ 
\multicolumn{1}{||l|}{$T_M^{\exp },$K} & \multicolumn{1}{c|}{325} & 
\multicolumn{1}{c|}{97.1} & \multicolumn{1}{c|}{94.5} & \multicolumn{1}{|c|}{
42.1} & 40.0 \\ \hline\hline
\end{tabular}
\]

In the easy-plane case ($\eta <0$) finite-temperature magnetic transition is
absent at $\alpha =0.$ At the same time, the Kosterlitz-Thouless transition
where unbinding of topological excitations (vortex pairs) occurs. For small
anisotropy value the temperature of this transition is small, and to leading
logarithmic accuracy \cite{Hikami} 
\begin{equation}
T_{KT}^{(0)}=\frac{4\pi |J|S^2}{\ln |\pi ^2/\eta |}  \label{TKTs}
\end{equation}
which is similar to the result for $T_M$ in the easy-axis case. As well as
for magnets with small easy-axis anisotropy, topological excitations are
important only close to $T_{KT}.$ Using the renormalization group approach,
the result similar to (\ref{I}) can be obtained \cite{OurXY} 
\begin{equation}
1=\frac{T_{KT}}{4\pi \rho _s}\left[ \ln \frac{\Gamma (T_{KT})}{|\,f_r|}+4\ln 
\frac{4\pi \rho _s}{T_{KT}}+C^{\prime }\right]   \label{tkt}
\end{equation}
For the magnetic ordering temperature in the presence of interlayer coupling
we obtain 
\begin{equation}
1=\frac{T_M}{4\pi \rho _s}\left[ \ln \frac{\Gamma (T_M)}{|f_r|}+4\ln \frac{%
4\pi \rho _s}{T_M}+C^{\prime }-\frac{2A^2}{\ln ^2(f/\alpha )}\right] 
\label{tc}
\end{equation}
The comparison of our results with the experimental data \cite{Joungh} is
presented in the Table 3 (the parameters $C^{\prime }\simeq -1.0$ and $A=3.5$
are fitted for the first compound).

\[
\begin{tabular}{||lc|cc||}
\hline\hline
\multicolumn{1}{||l|}{Compound} & \multicolumn{1}{c|}{K$_2$CuF$_4$} & 
\multicolumn{1}{c|}{stage-2 NiCl$_2$} & BaNi$_2$(PO$_4$)$_2$ \\ \hline\hline
\multicolumn{1}{||l|}{$T_{KT}^{\text{(0)}},\,$K} & \multicolumn{1}{c|}{11.4}
& \multicolumn{1}{c|}{35.3} & 45.0 \\ 
\multicolumn{1}{||l|}{$T_{KT}^{\text{RG}};T_C^{\text{RG}},$ K} & 
\multicolumn{1}{c|}{5.5; 6.25} & \multicolumn{1}{|c|}{17.4; 18.7} & 23.0;
24.3 \\ 
\multicolumn{1}{||l|}{$T_{KT}^{\exp };T_C^{\exp },$ K} & \multicolumn{1}{c|}{
5.5; 6.25} & \multicolumn{1}{c|}{18$\div $20} & 23.0; 24.5 \\ \hline\hline
\end{tabular}
\]
One can see that the two-loop corrections improve radically the result (\ref
{TKTs})$.$

{\sc Figure captions}

Fig.1. The theoretical temperature dependences of the relative sublattice
magnetization $\overline{\sigma }_r$ from the spin-wave theory (SWT), SSWT,
Tyablikov theory (TT), RG approach (\ref{fRG}), and the experimental points
for La$_2$CuO$_4$ \cite{Keimer}. The RG$^{\prime }$ curve corresponds to
inclusion of the function $\Phi _a^{\text{AF}}(t/\overline{\sigma })$ given
by (\ref{FF}). The $1/N$ curve is the critical behavior predicted by $1/N$%
-expansion (\ref{MagnCr1}). The result of $1/N$-expansion at intermediate
temperatures \cite{Our1/N} practically coincides with RG$^{\prime }.$

Fig.2. Temperature dependence of the relative sublattice magnetization $%
\overline{\sigma }(T)$ of K$_2$NiF$_4$ in the SWT, SSWT, RG approaches and $%
1/N$-expansion for $O(N)$ model as compared with the experimental data \cite
{Birg} (circles). The RG$^{\prime }$ curve corresponds to inclusion of the
function $\Phi _a^{\text{AF}}(t/\overline{\sigma })$ given by (\ref{FF}).
Short-dashed line is the extrapolation of the $1/N$-expansion result to the
critical region.

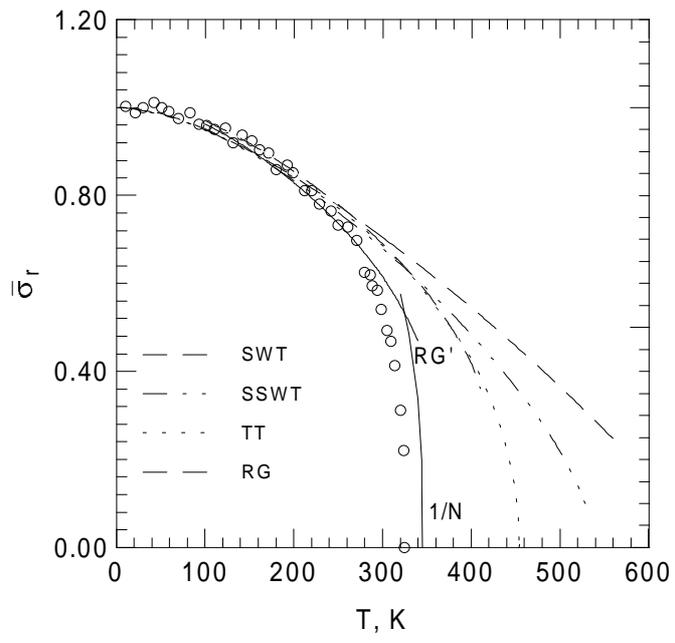
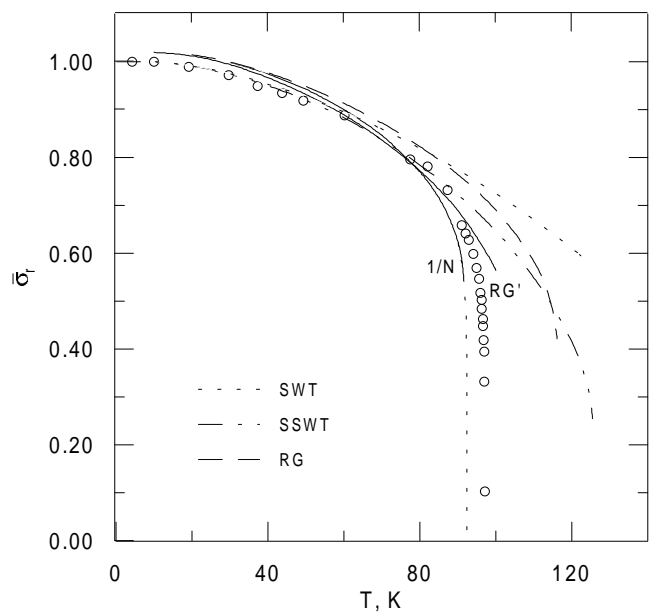
\end{document}